\documentclass[
    aip,
    apl,
    twocolumn,
    superscriptaddress,
    10pt,
]{revtex4-2}
\usepackage{graphicx}
\usepackage{amsmath}
\usepackage{siunitx}
\DeclareSIUnit\bar{bar}
\usepackage[
    colorlinks=true,
    linkcolor=blue,
    urlcolor=blue,
    citecolor=blue,
    anchorcolor=blue
]{hyperref}
\usepackage{makecell}

\begin{document}
\frenchspacing

\title{
Niobium Titanium Nitride as a High Tensile Stress Material for Nanomechanics
}

\author{Korbinian Rubenbauer}
\affiliation{Walther-Meißner-Institut, Bayerische Akademie der Wissenschaften, 85748 Garching, Germany}
\affiliation{Physics Department, TUM School of Natural Sciences, Technical University of Munich, 85748 Garching, Germany}
\author{Burak E. Bülbüloglu}
\affiliation{Walther-Meißner-Institut, Bayerische Akademie der Wissenschaften, 85748 Garching, Germany}
\affiliation{Physics Department, TUM School of Natural Sciences, Technical University of Munich, 85748 Garching, Germany}
\author{Thomas Luschmann}
\affiliation{Walther-Meißner-Institut, Bayerische Akademie der Wissenschaften, 85748 Garching, Germany}
\affiliation{Physics Department, TUM School of Natural Sciences, Technical University of Munich, 85748 Garching, Germany}
\affiliation{Munich Center for Quantum Science and Technology (MCQST), 80799 Munich, Germany}
\author{Manuel Müller}
\affiliation{Walther-Meißner-Institut, Bayerische Akademie der Wissenschaften, 85748 Garching, Germany}
\affiliation{Physics Department, TUM School of Natural Sciences, Technical University of Munich, 85748 Garching, Germany}
\author{Stephan Geprägs}
\affiliation{Walther-Meißner-Institut, Bayerische Akademie der Wissenschaften, 85748 Garching, Germany}
\author{Witlef Wieczorek}
\affiliation{Department of Microtechnology and Nanoscience (MC2), Chalmers University of Technology, Kemivägen 9, Gothenburg, SE-41296, Sweden}
\author{Eva M. Weig}
\affiliation{School of Computation, Information and Technology, Technical University of Munich, 85748 Garching, Germany}
\affiliation{TUM Center for Quantum Engineering (ZQE), 85748 Garching, Germany}
\affiliation{Munich Center for Quantum Science and Technology (MCQST), 80799 Munich, Germany}
\author{Matthias Althammer}
\affiliation{Walther-Meißner-Institut, Bayerische Akademie der Wissenschaften, 85748 Garching, Germany}
\affiliation{Physics Department, TUM School of Natural Sciences, Technical University of Munich, 85748 Garching, Germany}
\author{Hans Huebl}
\email{hans.huebl@wmi.badw.de}
\affiliation{Walther-Meißner-Institut, Bayerische Akademie der Wissenschaften, 85748 Garching, Germany}
\affiliation{Physics Department, TUM School of Natural Sciences, Technical University of Munich, 85748 Garching, Germany}
\affiliation{Munich Center for Quantum Science and Technology (MCQST), 80799 Munich, Germany}

\date{\today}

\begin{abstract}

Over the past decades, high-coherence mechanical resonators have been continuously pushed to new limits, using techniques such as dissipation dilution and clamp-tapering to enhance their quality factor beyond intrinsic material limitations. Today, these mechanical resonators are often fabricated from silicon-nitride, silicon-carbide or aluminum. Recently, however, interest in novel material platforms has grown, especially those allowing for the integration within superconducting circuits. Among these, superconducting nitrides stand out as particularly promising due to their high transition temperatures compared to elementary superconductors. Here, we introduce them as nanomechanical resonators and report on the fabrication and characterization of highly stressed, doubly clamped niobium titanium nitride ($\mathrm{NbTiN}$) nanostring resonators. Using optical interferometry, we determine the elastic properties and mechanical quality factor from room temperature to $\SI{13}{\kelvin}$. We observe high tensile stress up to $\SI{0.81}{\giga\pascal}$ along with a Young's modulus of $\SI{181}{\giga\pascal}$ and an intrinsic mechanical quality factor up to $850$. With these favorable mechanical properties, $\mathrm{NbTiN}$ constitutes a promising material platform for future applications in cavity electro- and nanomechanics.

\end{abstract}

\maketitle

Cavity electromechanics offers a mechanism by which the displacement of a mechanically compliant element is transduced into a shift of a microwave circuit’s resonance frequency.\cite{regalMeasuringNanomechanicalMotion2008,zhouSlowingAdvancingSwitching2013,aspelmeyerCavityOptomechanics2014,barzanjehOptomechanicsQuantumTechnologies2022} This dispersive interaction between a low-frequency mechanical oscillator and a GHz-scale microwave mode forms a versatile platform for precision sensing and quantum-state engineering. It has enabled achievements such as ground-state cooling,\cite{teufelSidebandCoolingMicromechanical2011,chanLaserCoolingNanomechanical2011,rossiMeasurementbasedQuantumControl2018} the preparation of squeezed mechanical states,\cite{wollmanQuantumSqueezingMotion2015,pirkkalainenSqueezingQuantumNoise2015,youssefiSqueezedMechanicalOscillator2023} entanglement generation,\cite{palomakiEntanglingMechanicalMotion2013,ockeloen-korppiStabilizedEntanglementMassive2018,mercierdelepinayQuantumMechanicsFree2021} and the sensing of minute forces.\cite{regalMeasuringNanomechanicalMotion2008,teufelNanomechanicalMotionMeasured2009,chasteNanomechanicalMassSensor2012,gavartinHybridOnchipOptomechanical2012,moserUltrasensitiveForceDetection2013,biswasFemtogramScalePhotothermalSpectroscopy2014,luschmannMechanicalFrequencyControl2022,dejongMeasurementStrongNonlinear2026,luschmannResolvingAbrikosovVortex2026} A key to the success of these systems, specifically in the context of quantum state generation, is the development of mechanical oscillators with high and ultra-high mechanical quality factors and coherence times. There, versatile tools like dissipation dilution, elastic strain engineering, soft clamping and phononic band structure engineering represent effective strategies to enhance the mechanical quality factor far above the intrinsic material properties to achieve record mechanical quality factors exceeding $10^{10}$.\cite{gonzalezBrownianMotionMass1994,unterreithmeierDampingNanomechanicalResonators2010,yuControlMaterialDamping2012,villanuevaEvidenceSurfaceLoss2014,tsaturyanUltracoherentNanomechanicalResonators2017,ghadimiElasticStrainEngineering2018,maccabeNanoacousticResonatorUltralong2020,sementilliNanomechanicalDissipationStrain2022,engelsenUltrahighqualityfactorMicroNanomechanical2024}

Typically, these platforms employ mechanical resonators with conducting or even superconducting properties, as these enable stronger electromechanical coupling rates.\cite{faustMicrowaveCavityenhancedTransduction2012,pernpeintnerCircuitElectromechanicsNonmetallized2014} In addition to electrical performance, mechanical parameters such as the mass density, Young’s modulus, and tensile stress are crucial for implementing dissipation‑dilution techniques. Due to its low density and well‑established fabrication processes, aluminum ($\mathrm{Al}$) has been the material of choice in most electromechanical platforms to date.\cite{regalMeasuringNanomechanicalMotion2008,hoehneDampingHighfrequencyMetallic2010,teufelCircuitCavityElectromechanics2011,teufelSidebandCoolingMicromechanical2011,pirkkalainenHybridCircuitCavity2013,wollmanQuantumSqueezingMotion2015,pirkkalainenSqueezingQuantumNoise2015,rodriguesCouplingMicrowavePhotons2019,schmidtSidebandresolvedResonatorElectromechanics2020,youssefiSqueezedMechanicalOscillator2023,beraSinglephotonInducedInstabilities2024,youssefiCompactSuperconductingVacuumgap2025,dejongMeasurementStrongNonlinear2026} Although it is not the optimal choice from a purely materials‑engineering standpoint, is has supported both nanostring‑based\cite{regalMeasuringNanomechanicalMotion2008,hoehneDampingHighfrequencyMetallic2010,rodriguesCouplingMicrowavePhotons2019,schmidtSidebandresolvedResonatorElectromechanics2020,beraSinglephotonInducedInstabilities2024,luschmannMechanicalFrequencyControl2022,dhimanSelfSustainedOscillations2026,luschmannResolvingAbrikosovVortex2026} and drum‑type\cite{teufelCircuitCavityElectromechanics2011,pirkkalainenHybridCircuitCavity2013,palomakiCoherentStateTransfer2013,wollmanQuantumSqueezingMotion2015,youssefiCompactSuperconductingVacuumgap2025} resonator implementations. However, the need for improved mechanical and superconducting properties, such as magnetic field resilience, has recently sparked interest in alternative, technologically compatible superconductors to realize mechanical resonators with ultra-high mechanical quality factors.

Superconducting nitrides are particularly attractive candidates to fulfill these requirements. They are known for their mechanical hardness\cite{torokYoungsModulusTiN1987,stoneHardnessElasticModulus1991,chenHardSuperconductingNitrides2005,arockiasamyDuctilityBehaviourCubic2016} and their elevated critical temperatures, which typically exceed those of elemental superconductors\cite{hornPraeparationUndSupraleitungseigenschaften1968,mullerMagneticFieldRobust2022} and thus suggest resilience against external magnetic fields, thermal fluctuations, and quasiparticle noise. In particular, $\mathrm{NbTiN}$ thin films with a critical temperature of up to $T_\mathrm{c} \approx \SI{16}{\kelvin}$\cite{yenSuperconductingHcJcTc1967, mullerMagneticFieldRobust2022} have already been employed as high-quality superconducting microwave resonators\cite{brunoReducingIntrinsicLoss2015,mullerMagneticFieldRobust2022} or as membrane resonators for the detection of the Casimir force.\cite{xuHighResolutionCasimirForce2026}  Yet, the mechanical properties of $\mathrm{NbTiN}$ in a nanostring geometry remain largely unexplored to our knowledge.

Here, we investigate the mechanical properties of doubly clamped $\mathrm{Nb_{0.7}Ti_{0.3}N}$ nanostring resonators in the temperature range between $\SI{13}{K}$ and $\SI{300}{K}$. We find that this material, which has been earlier investigated in the context of superconducting planar microwave resonators, \cite{mullerMagneticFieldRobust2022} shows strong tensile stress and a Young's modulus comparable to other nitride systems. These findings establish $\mathrm{NbTiN}$ as a promising material platform for future use in nanomechanical systems, specifically for the development of ultra-high quality mechanical resonators as used for sensing and superconducting cavity electromechanics.

\begin{figure}[t]
    \centering
    \includegraphics{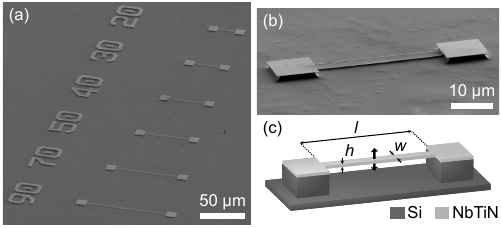}
    \caption{\textbf{Device Layout and Schematics.}
    (a) Scanning electron microscope (SEM) image of one of the examined series of nanostrings. The length of the nanostring increases from $\SI{20}{\micro\meter}$ to $\SI{90}{\micro\meter}$ as indicated by the number adjacent to each string.
    (b) SEM image of a $\SI{30}{\micro\meter}$ long, $\SI{500}{\nano\meter}$ wide, and $\SI{142}{\nano\meter}$ thin $\mathrm{NbTiN}$ suspended nanostring, which is clamped by two square pads with $\SI{10}{\micro\meter}$ side lengths.
    (c) Schematic of a nanostring, defining the string length $l$, width $w$, and thickness $h$.
    }
    \label{main:fig1-layout}
\end{figure}

The fabrication process starts with wet chemical cleaning of a high-resistive $\mathrm{(100)\text{-}}$silicon substrate. Then, the $\mathrm{NbTiN}$ thin film is deposited by reactive plasma sputtering in an argon ($\mathrm{Ar}$) and nitrogen ($\mathrm{N_2}$) atmosphere (flow ratio in SCCM of the process gases $\mathrm{Ar}$:$\mathrm{N_2} = \text{34:6}$) using a $\mathrm{Nb_{70} Ti_{30}}$ target. We choose a deposition pressure of $\SI{5}{\micro\bar}$ and a substrate temperature of $\SI{500}{\celsius}$. From this, we obtain a $\mathrm{NbTiN}$ thin film with a uniform thickness of $h = \SI{142}{\nano\meter}$ (determined by X-ray reflectometry, see Appendix\,\ref{si:sec-xrr}). The critical temperature of our blanket film is $\SI{16.3}{\kelvin}$ and it originates from the same batch investigated in Ref.\,\onlinecite{mullerMagneticFieldRobust2022}. For the patterning of the nanostring resonators, we apply a $\SI{30}{\nano\meter}$ thick aluminum hard mask using electron beam lithography, ion beam evaporation and a subsequent lift-off. By reactive ion etching (in $\mathrm{S F_{6}}$), we define and release the doubly clamped $\mathrm{NbTiN}$ nanostrings. Last, the aluminum hard mask is removed by wet chemical etching (in TechniEtch Al80), followed by critical-point-drying. Figure\,\ref{main:fig1-layout} shows an exemplary $\mathrm{NbTiN}$ nanostring, which is part of a series of resonators with varying lengths $l = (20,30,40,50,70,90) \, \si{\micro\meter}$ and a constant width $w = \SI{500}{\nano\meter}$. In this study, we analyze multiple of these series to obtain average values for each length.

\begin{figure}[t]
    \centering
    \includegraphics{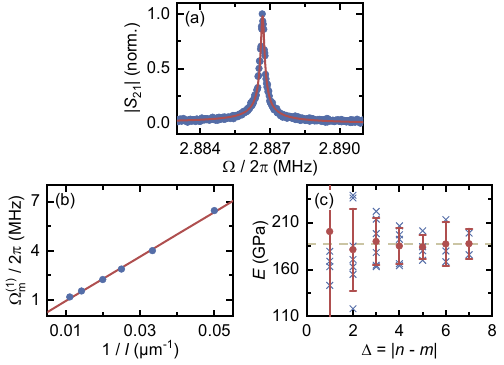}
    \caption{\textbf{Room Temperature Characterization.}
    (a) Optical spectroscopy measurement. A typical normalized spectrum (blue dots) of a $\SI{40}{\micro\meter}$ long string is fitted with Eq.\,\eqref{main:eq-lorentz-fit}. From the fit, we extract a resonance frequency of $\Omega_\mathrm{m}^{(1)} / 2 \pi = \SI{2.886 687}{\mega\hertz}$ and linewidth of $\Gamma_\mathrm{m}^{(1)} / 2 \pi = \SI{154}{\hertz}$.
    (b) Inverse length ($1/l$) dependence of the first harmonic $\Omega_\mathrm{m}^{(1)}$. Each point is an average over data obtained for multiple strings of the same length. The standard deviation is within the point size. Using the high tensile stress approximation of Eq.\,\eqref{main:eq-res-freq-length}, we obtain a tensile stress of $\sigma = \left( 0.56 \pm 0.03 \right) \si{\giga\pascal}$.
    (c) Young's modulus $E$ from the higher harmonics of an exemplary $\SI{90}{\micro\meter}$ long string as a function of the mode index difference $\Delta$. From the highest available $\Delta = 7$ we find $E = \left(188 \pm 17\right) \si{\giga\pascal}$, which is shown by the yellow dashed line.}
    \label{main:fig2-rt}
\end{figure}

The mechanical response to an oscillating force provided by a piezoelectric actuator is determined by optical interferometry, where monochromatic laser light (wavelength $\SI{633}{nm}$) is used to monitor the displacement of the mechanical oscillator. To this end, the laser is focused on a selected nanostring, where it is partially reflected. The detected light becomes intensity-modulated, encoding the displacement of the nanostring. We convert this optical signal into an electrical signal using a balanced photodetector.\cite{yuControlMaterialDamping2012,ghadimiElasticStrainEngineering2018,pernpeintnerFrequencyControlCoherent2018,kinimanjeshwarSuspendedPhotonicCrystal2020,buckleUniversalLengthDependence2021,klassDeterminingYoungsModulus2022,ciersNanomechanicalCrystallineAlN2024,hochreiterMonolithic4H$mathrmSimathrmC$Nanomechanical2025} We acquire the data using a vector network analyzer (VNA), which provides the rf voltage sourcing the piezoelectric actuator and detects the mechanical response via the photodetector. The resulting measured quantity is the complex effective transmission parameter $S_{21}$ which contains the mechanical response of the nanostring oscillator (see Fig.\,\ref{main:fig2-rt}\,(a)). We analyze this data by fitting the response with
\begin{equation}
        \left| S_{21} \left( \Omega \right) \right| = \left| \frac{a}{\left( \Omega_\mathrm{m}^{(n)} - \Omega \right) - i \Gamma_\mathrm{m}^{(n)}/2} + c_1 + i c_2 \right| .
        \label{main:eq-lorentz-fit}
\end{equation}
Here, $\Omega_\mathrm{m}^{(n)}$ is the mechanical resonance frequency of the mode with index $n$ ($n=1$: fundamental mode), $\Gamma_\mathrm{m}^{(n)}$ is the corresponding FWHM linewidth, $a$ is an amplitude response factor, and $c_1$ as well as $c_2$ represent a real and an imaginary background. The introduction of the latter allows to correct for a broad background response, potentially arising from mechanical modes of the chip or direct radio-frequency cross-talk, which can cause a Fano-type lineshape. When accessible, we investigate all modes from the fundamental up to the $10$-th harmonic. The sample is mounted in a commercial dry cryostat with optical access ports, which allows to control the sample space temperature between room temperature ($\sim \SI{293}{\kelvin}$) and $\sim \SI{13}{\kelvin}$. Lower temperatures could not be achieved due to heating caused by the light entering through the optical access ports. All experiments have been performed in the linear response regime and at pressures $< 5 \times 10^{-6} \, \si{\milli\bar}$ to suppress air damping.

Figure\,\ref{main:fig2-rt}\,(a) shows a typical normalized mechanical response spectrum obtained for the fundamental mode of a $\SI{40}{\micro\meter}$ long $\mathrm{NbTiN}$ nanostring at $T \approx \SI{293}{\kelvin}$. We observe a clear mechanical signature in $\left|S_{21}\right|$, with a peak response at the mechanical resonance frequency of $\Omega_\mathrm{m}^{(1)} / 2 \pi = \SI{2.886 687}{\mega\hertz}$ and a mechanical loss rate of $\Gamma_\mathrm{m}^{(1)} / 2 \pi = \SI{154}{\hertz}$, which we obtain by fitting Eq.\,\eqref{main:eq-lorentz-fit} to the data. The mechanical quality factor of this mode is $Q_\mathrm{m}^{(1)} = \Omega_\mathrm{m}^{(1)} / \Gamma_\mathrm{m}^{(1)} = 18.7 \times 10^3$. A detailed discussion about the mechanical quality factor will follow later. 

To determine the tensile stress $\sigma$, we analyze the mechanical resonance frequency as function of the string length $l$. The Young's modulus $E$ of the material is found by comparing the resonance frequencies of higher order modes of one and the same string. Generally, the mechanical resonance frequency of a tensile stressed beam with simply supported ends is given by\cite{TimoshenkoVibrationProblemsEngineering1991,clelandFoundationsNanomechanics2003,verbridgeHighQualityFactor2006,hockeDeterminationEffectiveMechanical2014,klassDeterminingYoungsModulus2022}
\begin{equation}
    \Omega_\mathrm{m}^{(n)} = \frac{n^2 \pi^2}{l^2} \sqrt{\frac{E I}{\rho A} \left( 1 + \frac{\sigma A l^2}{n^2 \pi^2 E I} \right)} \approx \frac{n \pi}{l} \sqrt{\frac{\sigma}{\rho}} .
    \label{main:eq-res-freq-length}
\end{equation}
Here, $\rho$ the mass density, $I$ is the moment of inertia, and $A$ the cross-sectional area of the nanostring. For the out-of-plane mode considered here, $I= wh^3/12$ and $A=w\cdot h$. The approximation on the right hand side assumes the regime of high tensile stress, for which $\sigma\gg E I \pi^2/ A l^2$.\cite{verbridgeHighQualityFactor2006,hockeDeterminationEffectiveMechanical2014} 

For the initial determination of the tensile stress, we assume the high tensile stress regime and analyze the length dependence of the fundamental resonance frequency $\Omega_\mathrm{m}^{(1)}$ shown in Fig.\,\ref{main:fig2-rt}\,(b). Using the material density of our $\mathrm{NbTiN}$ of $\rho = \SI{7.6}{\gram\per\centi\meter^3}$, which we determine by X-ray reflectometry (see Appendix\,\ref{si:sec-xrr}), we obtain a tensile stress of $\sigma = \left( 0.56 \pm 0.03 \right) \si{\giga\pascal}$ at room temperature. This method, however, does not account for the length dependence of the tensile stress reported in Ref.\,\onlinecite{buckleUniversalLengthDependence2021}, which becomes especially relevant for string lengths $\leq \SI{100}{\micro\meter}$ and thus for all lengths examined in this study. In this context, the tensile stress obtained here is an average over all lengths, whereas the actual stress is larger for shorter strings and smaller for longer strings. This is discussed in more detail in Appendix\,\ref{si:sec-sigma-l-dep}.

The resonance frequencies of the higher order mechanical modes allow to determine the Young's modulus $E$. Essentially, Equation\,\eqref{main:eq-res-freq-length} indicates that the resonance frequencies of higher harmonics do not scale perfectly linearly with the harmonic order. Ref.\,\onlinecite{klassDeterminingYoungsModulus2022} recasts Eq.\,\eqref{main:eq-res-freq-length} to obtain the Young's modulus $E$ from the frequency difference between two different harmonics $n$ and $m$ via
\begin{equation}
    E = \frac{l^4 \rho A}{\pi^4 I \left( n^2 - m^2 \right)} \left( \left(\frac{\Omega^{(n)}_\mathrm{m}}{n}\right)^2 - \left(\frac{\Omega^{(m)}_\mathrm{m}}{m}\right)^2 \right) .
    \label{main:eq-youngs-modulus}
\end{equation}
Notably, analyzing the data in this way is independent of the tensile stress, and only requires the material density and the geometry of the nanostring as input parameters. Experimentally, we determine the frequencies of multiple harmonics of various strings using the optical interferometry setup introduced before and calculate $E$ with Eq.\,\eqref{main:eq-youngs-modulus} for each available pair of harmonics. 

Figure\,\ref{main:fig2-rt}\,(c) shows the Young's modulus $E$ as a function of the mode index difference $\Delta = \left| n - m \right|$, exemplarily for a string of $\SI{90}{\micro\meter}$ length. In this case, a maximum harmonics difference of $\Delta = 7$ was observed experimentally. In accordance with the existing literature, the variance in the determined Young's modulus decreases as $\Delta$ increases.\cite{klassDeterminingYoungsModulus2022} From this, we extract $E = \left(188 \pm 17\right) \si{\giga\pascal}$. As before, this procedure is repeated for multiple strings, in this case strings of $\SI{70}{\micro\meter}$ and $\SI{90}{\micro\meter}$ length, to find an average Young's modulus of $E = \left( 183 \pm 11 \right) \si{\giga\pascal}$ at room temperature. By considering the out-of-plane mode and the shortest measured string ($l = \SI{20}{\micro\meter}$), we confirm our previous assumption of high tensile stress as $\SI{0.56}{\giga\pascal} = \sigma \gg E I \pi^2/ A l^2 = \SI{0.008}{\giga\pascal}$. In detail, we obtain a tensile stress of $\sigma = \left( 0.56 \pm 0.03 \right) \si{\giga\pascal}$ for the high tensile stress assumption in comparison to $\sigma = \left( 0.55 \pm 0.03 \right) \si{\giga\pascal}$ for the full model, which is in agreement within the error margin.

To assess the potential of $\mathrm{NbTiN}$ for nanomechanical elements in cavity electromechanics, we study the temperature dependence of the tensile stress and the Young's modulus. We expect that, at all temperatures investigated, the $\mathrm{NbTiN}$ nanostring remains in the normal conducting state as our readout concept employs optical light.\cite{strohauerSiteSelectiveEnhancementSuperconducting2023} Furthermore, the critical temperature of $\SI{16}{\kelvin}$ reported in Ref.\,\onlinecite{mullerMagneticFieldRobust2022} was measured for broad mircrowave transmission lines, whereas it is expected to be lower for the thin nanostring geometry examined here.

\begin{figure}[t]
    \centering
    \includegraphics{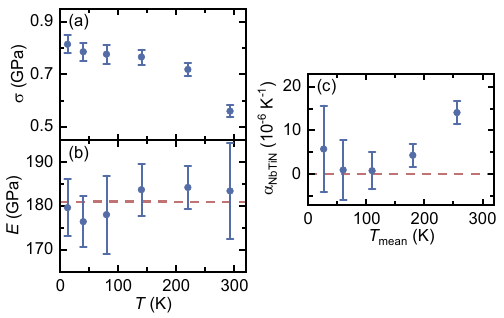}
    \caption{\textbf{Material Parameter Temperature Dependence.}
    (a) Tensile stress $\sigma$ as a function of the temperature $T$ from room temperature to $\SI{13}{\kelvin}$. The error is given by the fit error.
    (b) Young's modulus $E$ as a function of the temperature $T$. We show the average between multiple strings, with the error given by the standard deviation. The red dashed line displays the average over all temperatures of $E = \SI{181}{\giga\pascal}$.
    (c) Thermal expansion coefficient $\alpha_\mathrm{NbTiN}$ determined from the change in tensile stress as a function of the temperature $T$. The error is propagated from the error in the tensile stress and the Young's modulus. The red dashed line separates positive and negative thermal expansion coefficients.
    }
    \label{main:fig3-Tdep-sigma-E}
\end{figure}

Figure\,\ref{main:fig3-Tdep-sigma-E}\,(a) shows the temperature dependence of the tensile stress under the assumption of the high tensile stress regime. We observe a $\SI{45}{\percent}$ increase of the tensile stress from $\SI{0.56}{\giga\pascal}$ at room temperature to $\SI{0.81}{\giga\pascal}$ at $\SI{13}{\kelvin}$. Most of the increase, specifically $\SI{82}{\percent}$, already occurs in the temperature range between $\SI{293}{\kelvin}$ and $\SI{140}{\kelvin}$. We attribute this increase to a difference between the thermal expansion coefficients of the $\mathrm{NbTiN}$ and the $\mathrm{Si}$ substrate. In contrast, as shown in Fig.\,\ref{main:fig3-Tdep-sigma-E}\,(b), the Young's modulus is effectively temperature independent within the error bars. The average value over all temperatures is $\SI{181}{\giga\pascal}$.

Assuming, that the observed tensile stress change $\Delta \sigma$ originates solely from the difference between the thermal expansion coefficients $\alpha_\mathrm{Si}$ and $\alpha_\mathrm{NbTiN}$, we can extract $\alpha_\mathrm{NbTiN}$ via\cite{callisterMaterialsScienceEngineering}
\begin{equation}
    \alpha_\mathrm{NbTiN} (T_\mathrm{mean}) = \bar{\alpha}_\mathrm{Si} (T_\mathrm{mean}) - \frac{\Delta \sigma}{E \, \Delta T} .
    \label{main:eq-thermal-exp-coeff}
\end{equation}
To obtain the temperature dependence of $\alpha_\mathrm{NbTiN}$, \linebreak$\Delta\sigma = \sigma_\mathrm{f} - \sigma_0$ and $\Delta T = T_\mathrm{f} - T_0$ are calculated using two adjacent temperatures $T_\mathrm{f}$ and $T_0$. We further assume a constant Young's modulus of $\SI{181}{\giga\pascal}$ for all temperatures. For $\bar{\alpha}_\mathrm{Si} (T_\mathrm{mean})$, the average of $\alpha_\mathrm{Si} (T)$ is taken in the temperature interval $[T_\mathrm{f},T_0]$ using the values for $\alpha_\mathrm{Si} (T)$ from Ref.\,\onlinecite{gibbonsThermalExpansionCrystals1958}. We assign the value of $\alpha_\mathrm{NbTiN}$ obtained in this manner to the mean temperature $T_\mathrm{mean} = (T_\mathrm{f} + T_0)/2$ of the two neighboring points. Figure\,\ref{main:fig3-Tdep-sigma-E}\,(c) shows the resulting thermal expansion coefficients as a function of $T_\mathrm{mean}$. The observed $\alpha_\mathrm{NbTiN}$ decreases from room temperature to around $\SI{100}{\kelvin}$ and then increases again for lower temperatures. They are of the order of $10^{-6}\,\si{\kelvin^{-1}}$, which is within the typical range for a metal at these temperatures.\cite{grossFestkorperphysik2022} At the highest mean temperature of $\SI{257}{\kelvin}$, we find a thermal expansion coefficient of $14 \times 10^{-6} \,\si{\kelvin^{-1}}$. This is comparable to the thermal expansion coefficients of niobium ($\alpha_\mathrm{Nb} = 7.1 \times 10^{-6} \,\si{\kelvin^{-1}}$) and titanium-niobium ($\alpha_\mathrm{Ti\text{-}Nb} = 9.0 \times 10^{-6} \,\si{\kelvin^{-1}}$) at room temperature.\cite{barronHeatCapacityThermal1999} We are not aware of any works which have previously determined the thermal expansion coefficient of $\mathrm{NbTiN}$. 

\begin{figure}[t]
    \centering
    \includegraphics{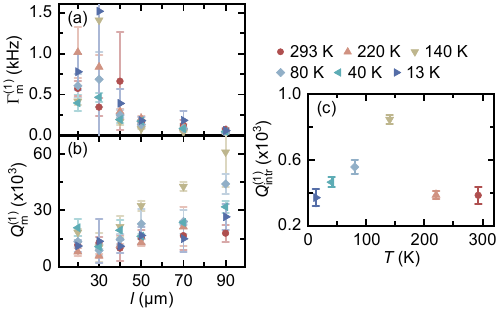}
    \caption{\textbf{Mechanical linewidth and (intrinsic) quality factors.}
    (a) Length dependence of the mechanical linewidth $\Gamma_\mathrm{m}$. Average values between different strings of the same length are displayed together with their standard deviation as error bars. An decrease of the loss rate with increasing length is observed.
    (b) Mechanical quality factor $Q_\mathrm{m} = \Omega_\mathrm{m} / \Gamma_\mathrm{m}$ as a function of the string length. Average values between different strings of the same length are shown together with their standard deviation as error bars. We observe a roughly linear increase of the mechanical quality with length.
    (c) Intrinsic quality factor $Q_\mathrm{intr}$ as a function of temperature $T$, obtained by fitting Eq.\,\eqref{main:eq-Q-intr} to $Q_\mathrm{m} (l)$. The error bars depict the error of the fit.
    }
    \label{main:fig4-loss-Q-vs-T}
\end{figure}

Next, we determine the intrinsic quality factor $Q_\mathrm{intr}^{(1)}$, which reflects the intrinsic loss mechanisms of the material. Notably, these include losses associated with the surface of the material as well as its bulk. For the investigated thickness of $\SI{142}{\nano\meter}$, the losses are most likely dominated by the surface.\cite{villanuevaEvidenceSurfaceLoss2014} Assuming that the mechanical quality factor $Q_\mathrm{m}^{(1)}$ is enhanced by dissipation dilution, $Q_\mathrm{intr}^{(1)}$ can be determined from the length dependence of the mechanical quality factor $Q_\mathrm{m}^{(1)}\left(l\right)$ via\cite{schmidFundamentalsNanomechanicalResonators2023}
\begin{equation}
    Q_\mathrm{m}^{(1)} = Q_\mathrm{intr}^{(1)} / \left( \pi^2 \lambda^2 + 2\lambda \right) .
    \label{main:eq-Q-intr}
\end{equation}
Here, $\lambda = \frac{h}{l} \sqrt{\frac{E}{12\sigma}}$ is the strain parameter and enhances the mechanical quality factor above the intrinsic quality factor for $\lambda \ll 1$. This is the case for all temperatures and lengths investigated here. \\

The length dependence of the dissipation rate and the quality factor are depicted in Fig.\,\ref{main:fig4-loss-Q-vs-T}\,(a) and (b). For all temperatures, we observe a decrease of $\Gamma_\mathrm{m}^{(1)}$ for longer strings. Similar to that, $Q_\mathrm{m}^{(1)}$ shows the characteristic signature of a roughly linear increase with the string length $l$ expected from Eq.\,\eqref{main:eq-Q-intr} for the examined lengths. It reaches values from $10\,000$ up to $60\,000$. We fit $Q_\mathrm{m}^{(1)} \left( l \right)$ for each measured temperature individually with Eq.\,\eqref{main:eq-Q-intr} to obtain $Q_\mathrm{intr}^{(1)}$ at the given temperature. For the fit, we use the geometric dimensions of the nanostring, the tensile stress at the given temperature and the temperature averaged Young's modulus of $\SI{181}{\giga\pascal}$ as fixed parameters. Figure\,\ref{main:fig4-loss-Q-vs-T}\,(c) shows the resulting temperature dependence of the intrinsic quality factor. The data suggests an increase of $Q_\mathrm{intr}^{(1)}$ from room temperature to $\SI{140}{\kelvin}$, where it reaches its maximum of $(0.85 \pm 0.03) \times 10^{3}$, followed by a gradual decrease for lower temperatures. This behavior differs from the naively expected improvement of $Q_\mathrm{intr}^{(1)}$ towards lower temperatures and could, even though this would only be expected for thicker films, hint to the presence of thermo-elastic damping as a large $Q_\mathrm{intr}^{(1)}$ seems to correlate with a small thermal expansion coefficient $\alpha_\mathrm{NbTiN}$.\cite{lifshitzThermoelasticDampingMicro2000} However, the identification of the limiting damping mechanisms is subject to further studies.

\begin{table}[b]
\caption{\textbf{Comparison to other materials.}
Our $\mathrm{NbTiN}$ is compared to other materials in nanomechanics in terms of the tensile stress $\sigma$, the Young's modulus $E$, the intrinsic quality factor $Q_\mathrm{intr}$, the film thickness $h$, and the temperature $T$ at which these values were obtained ($\text{RT}$ = room temperature).
}
\label{main:tab-compare}
\begin{ruledtabular}
\begin{tabular}{c|ccccc}
Material & \makecell{$\sigma$ \\ $(\si{\giga\pascal})$} & \makecell{$E$ \\ $(\si{\giga\pascal})$} & \makecell{$Q_\mathrm{intr}^{(1)}$ \\ $(\times 10^3)$} & \makecell{$h$ \\ $(\si{\nano\meter})$} & \makecell{$T$ \\ $(\si{\kelvin})$} \\ \hline
this work                                                                                              & $0.81$  & $181$ & $0.5$ & $142$                & $13$ \\
$\mathrm{Nb_{70}Ti_{30}N}$\cite{xuHighResolutionCasimirForce2026}                                      & $0.68$  & $-$   & $2.2$ & $155$                & $4.45$ \\
$\mathrm{TiN}$\cite{matsuyamaHighQMembraneResonators2025}                                              & $2.3$   & $200$ & $11$  & $100$                & $2.2$ \\
$\mathrm{Al}$\cite{youssefiCompactSuperconductingVacuumgap2025}                                        & $0.35$  & $69$  & $400$ & $150\,\text{-}\,200$ & $0.01$ \\
$\mathrm{AlN}$\cite{ciersNanomechanicalCrystallineAlN2024}                                             & $1.43$  & $270$ & $8$   & $290$                & $\text{RT}$ \\
$\mathrm{Nb}$-$\mathrm{Si_3 N_4}$\cite{hockeDeterminationEffectiveMechanical2014}                      & $0.199$ & $100$ & $4.6$ & $130+70$             & $0.4$ \\
$\mathrm{Si_3 N_4}$\cite{ghadimiElasticStrainEngineering2018,bereyhiClampTaperingIncreasesQuality2019} & $1.14$  & $250$ & $6.9$ & $100$                & $\text{RT}$ \\
$a\text{-}\mathrm{SiC}$\cite{xuHighStrengthAmorphousSilicon2024}                                       & $0.76$  & $223$ & $5.1$ & $71$                 & $\text{RT}$ \\
$4H\text{-}\mathrm{SiC}$\cite{hochreiterMonolithic4H$mathrmSimathrmC$Nanomechanical2025}               & $<0.01$ & $437$ & $140$ & $580$                & $\text{RT}$ \\
$\mathrm{In_{0.43}Ga_{0.57}P}$\cite{manjeshwarHighQTrampolineResonators2023}                           & $0.47$  & $108$\cite{klassDeterminingYoungsModulus2022} & $8$ & $73$                & $\text{RT}$
\end{tabular}
\end{ruledtabular}
\end{table}

Finally, Table\,\ref{main:tab-compare} compares the elastic and the mechanical properties of our $\mathrm{NbTiN}$ to other materials often employed in cavity opto- and electromechanics. In particular, we list the tensile stress $\sigma$, the Young's modulus $E$ and the intrinsic quality factor $Q_\mathrm{intr}^{(1)}$ for the various materials. $\mathrm{NbTiN}$ clearly outperforms aluminum, which is widely used in cavity electromechanics, as it has both a more than two times higher tensile stress and a more than two times higher Young's modulus. The large tensile stress is particularly favorable when considering dissipation dilution to further enhance the mechanical quality factor. The $Q_\mathrm{intr}^{(1)}$ reported here is large for a (super)conducting system, especially since the quality factor typically decreases as the metal layer thickness increases.\cite{yuControlMaterialDamping2012,seitnerDampingMetallizedBilayer2014} However, it is still lower by up to an order of magnitude compared to insulating materials like $\mathrm{Si_3 N_4}$ or $\mathrm{SiC}$, which are insulating and thus have less loss channels. Since $Q_\mathrm{intr}^{(1)}$ ultimately limits the application potential of $\mathrm{NbTiN}$ in nanomechanics and cavity electromechanics, future studies should examine $Q_\mathrm{intr}^{(1)}$ as a function of the film thickness as well as its dependence on surface treatments and growth parameters to open up avenues for improvements of $Q_\mathrm{intr}^{(1)}$. Here, as long as the superconductivity of the $\mathrm{NbTiN}$ is conserved, e.g.~thinner films can enable higher $Q_\mathrm{m}$ as the strain parameter $\lambda$ increases with the aspect ratio $h / l$.\cite{ciersThicknessDependenceMechanical2024}

As many cavity electromechanical platforms rely on a change in the separation of two electrodes, which transduces the mechanical displacement to a frequency shift of the electrical resonator, the conductive and in particular superconducting properties of the material are important as well. Of the materials listed in Table\,\ref{main:tab-compare}, only $\mathrm{Al}$, $\mathrm{TiN}$ and $\mathrm{NbTiN}$ are superconductors at low temperatures. While $\mathrm{Al}$ is established as the workhorse of superconducting quantum circuits, $\mathrm{NbTiN}$ recently also demonstrated excellent performance when used for microwave resonators. Notably, lumped element superconducting resonators fabricated from the same tensile-stressed film reach internal microwave quality factors of up to $0.2 \times 10^{6}$ at the single-photon level at millikelvin temperatures. Moreover, this performance is preserved at elevated magnetic fields up to $\sim \SI{130}{\milli\tesla}$.\cite{mullerMagneticFieldRobust2022}

Alternative to nanostrings made from conductive materials, concepts leveraging the change of the dielectric properties originating from mechanical motion exist, which allow to use insulating materials.\cite{unterreithmeierUniversalTransductionScheme2009,faustMicrowaveCavityenhancedTransduction2012,pernpeintnerCircuitElectromechanicsNonmetallized2014,schmidFundamentalsNanomechanicalResonators2023}. $\mathrm{NbTiN}$ is particularly interesting in this context because, unlike in this work, it can also be deposited in an insulating, nitrogen-rich phase.\cite{burdastyhSuperconductorInsulatorTransition2017,mironovLightinducedCurrentCooper2025} However, these concepts based on insulating materials typically have a considerably smaller electromechanical coupling rate compared to their conducting counterparts.\cite{pernpeintnerCircuitElectromechanicsNonmetallized2014} This makes the advent of high tensile stress and high Young's modulus materials such as $\mathrm{NbTiN}$ attractive.

In conclusion, $\mathrm{NbTiN}$ constitutes a material platform with potential for superconductor-based cavity electro- and nanomechanical systems. Its mechanical (and microwave) properties clearly outperform those of aluminum, making it well-suited for dissipation dilution schemes targeting ultra-high mechanical quality factors. While the intrinsic quality factors reported here remain below those attained in mature insulating platforms such as $\mathrm{Si_3N_4}$ and $\mathrm{SiC}$, there is considerable scope for further optimization. Beyond the normal-conducting regime explored here, the mechanical behavior in the superconducting state remains an open question that needs to be addressed in future studies. A natural next step is therefore the integration of these nanostrings into a cavity-electromechanical architecture, which would allow the characterization of the mechanical response in the superconducting phase excluding the influence from an optical read-out.

\section*{Supplementary Material}
See supplementary material for the X-ray reflectometry (XRR) data evaluation used to determine the density and film thickness, as well as a discussion of the length dependence of the tensile stress.

\section*{Author Declarations}

\section*{Conflict of Interest}
The authors have no conflicts to disclose.

\section*{Acknowledgements}
All authors acknowledge funding from the Horizon Europe 2021–2027 Framework Programme under the Grant Agreement No. 101080143 (SuperMeQ) and disclose support from the Deutsche Forschungsgemeinschaft (DFG, German Research Foundation) under Germany’s Excellence Strategy–EXC-2111-390814868. HH and EMW acknowledges funding from the Munich Quantum Valley, which is supported by the Bavarian state government with funds from the Hightech Agenda Bayern Plus. WW acknowledges support by the Knut and Alice Wallenberg (KAW) Foundation through a Wallenberg Academy Scholar.

\section*{Data Availability}
The raw data that supports the findings of this study is available from the corresponding author upon reasonable request. Processed data representing all the data in the published figures, both from the main text and the appendix, are openly available in Zenodo under at \url{ https://doi.org/10.5281/zenodo.21332861}, reference number \onlinecite{zenodo-archive}.

\bibliography{bibliography.bib}

\clearpage

\onecolumngrid
\appendix

\section{X-Ray Reflectometry}
\label{si:sec-xrr}

\setcounter{figure}{0}
\renewcommand{\thefigure}{A\arabic{figure}}
\renewcommand{\thetable}{A\arabic{table}}

\begin{figure}[h]
    \centering
    \includegraphics{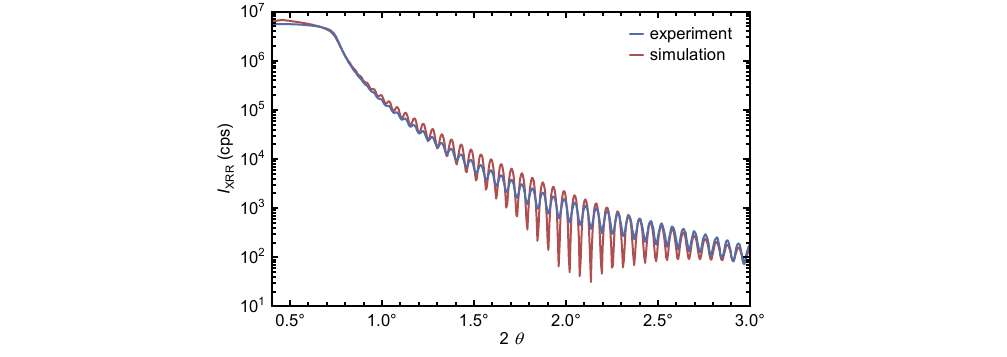}
    \caption{\textbf{XRR measurement and simulation.}
    The intensity $I_\mathrm{XRR}$ is displayed in counts per second (cps) as a function of the grazing-incidence angle $2 \theta$. The experimental and simulated data are depicted as a blue and red line, respectively.}
    \label{si:fig-xrd}
\end{figure}

This appendix discusses the determination of the film thickness $h$ and the density $\rho$ using X-ray reflectometry measurements (XRR).\cite{pietschHighResolutionXRayScattering2004} These were performed on a separate $6 \times 10 \, \si{\milli\meter\squared}$ $\mathrm{NbTiN}$ thin film sample, which was sputtered in the same run as the sample with the nanostrings. We simulate the intensity of the $\mathrm{NbTiN/Si}$ chip and compare it to the measured intensity $I_\mathrm{XRR}$ as a function of the grazing-incidence angle $2 \theta$ in order to obtain the film thickness $h$, the surface roughness $R$, and the $\mathrm{NbTiN}$ density $\rho$. The film thickness and the material density are important input parameters in the formulas used to obtain the tensile stress and the Young's modulus. Figure\,\ref{si:fig-xrd} displays the corresponding XRR measurement and simulation. From the simulation, we find a film thickness of $h = \left(141.5 \pm 0.2\right) \,\si{\nano\meter}$, a surface roughness of $R = \left(1.2 \pm 0.1\right) \,\si{\nano\meter}$ and a material density of $\rho = \left(7.60 \pm 0.05\right) \,\si{\gram\per\centi\meter^3}$.

\section{Length Dependence of the Tensile Stress}
\label{si:sec-sigma-l-dep}

\setcounter{figure}{0}
\renewcommand{\thefigure}{B\arabic{figure}}
\renewcommand{\thetable}{B\arabic{table}}

In this appendix, we present and discuss the dependence of the tensile stress $\sigma$ on the string length $l$. From the higher harmonics of a nanostring, one can not only determine the Young's modulus $E$ but also the tensile stress $\sigma$. Following the approach of Ref.\,\onlinecite{buckleUniversalLengthDependence2021}, we fit the resonance frequency $\Omega_\mathrm{m}^{(n)}$ as a function of the mode number $n$ with
\begin{equation}
    \Omega_\mathrm{m}^{(n)} = \left(\frac{n \pi}{l}\right)^2 \sqrt{\frac{E h^2}{12 \rho}} \sqrt{1 + \frac{12 \sigma l^2}{n^2 \pi^2 E h^2}} .
    \label{si:eq-f-vs-harm}
\end{equation}
For the fit, we fix the Young's modulus to the temperature-averaged value of $E = \SI{181}{\giga\pascal}$, the string length to its nominal design value, the thickness to $h = \SI{142}{\nano\meter}$, and the density to $\rho = \SI{7.6}{\gram\per\centi\meter^3}$, leaving the tensile stress $\sigma$ as the only free fit parameter.

\begin{figure}[ht]
    \centering
    \includegraphics{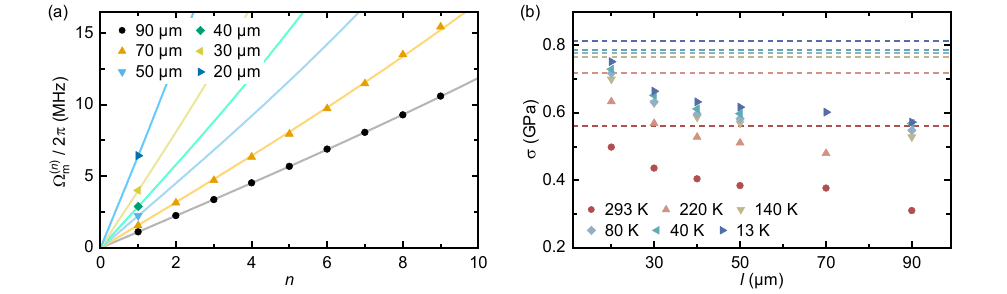}
    \caption{\textbf{Frequency of higher harmonics and length-dependent tensile stress.}
    (a) Resonance frequency $\Omega_\mathrm{m}^{(n)}$ as a function of the harmonic order $n$ at room temperature ($T = \SI{293}{\kelvin}$). Different string lengths are displayed in different colors and marker symbols. The markers denote the measured data, whereas the corresponding solid lines depict a fit with Eq.\,\eqref{si:eq-f-vs-harm}, from which the tensile stress is extracted.
    (b) Tensile stress $\sigma$ obtained from higher harmonics as a function of string length $l$. Only the $\SI{70}{\micro\meter}$ and $\SI{90}{\micro\meter}$ long strings were measured with a sufficient number of higher harmonics. The values for all shorter strings are estimated from the fundamental mode ($n = 1$) alone, which gives rise to the discontinuity between $\SI{50}{\micro\meter}$ and $\SI{70}{\micro\meter}$. Dashed lines indicate the tensile stress obtained from the fit of $\Omega_\mathrm{m}^{(1)} (l)$ as described in the main text.
    }
    \label{si:fig-f-vs-harm-tensile}
\end{figure}

At each measured temperature, we average the resonance frequency of every available harmonic over all examined strings of a given length and fit the result with Eq.\,\eqref{si:eq-f-vs-harm}. This is shown in Fig.\,\ref{si:fig-f-vs-harm-tensile}\,(a) for room temperature. Since in the experiment higher harmonics were only resolved for $\SI{70}{\micro\meter}$ and $\SI{90}{\micro\meter}$ long strings, only a single data point is available for all shorter strings, namely the fundamental mode $n = 1$. Consequently, the corresponding stress values should be regarded as rough estimates that likely underestimate the true stress at the given length. From the fits at room temperature displayed in Fig.\,\ref{si:fig-f-vs-harm-tensile}\,(a), we obtain a tensile stress of $\sigma_\mathrm{90\si{\micro\meter}} = \left( 0.310 \pm 0.001 \right) \si{\giga\pascal}$ for $l = \SI{90}{\micro\meter}$, and $\sigma_\mathrm{70\si{\micro\meter}} = \left( 0.377 \pm 0.004 \right) \si{\giga\pascal}$ for $l = \SI{70}{\micro\meter}$. Both values are notably lower than the tensile stress of $\left( 0.56 \pm 0.03 \right) \si{\giga\pascal}$ obtained from the length dependence of the fundamental mode $\Omega_\mathrm{m}^{(1)} (l)$ in the main text. Even for the shortest examined string ($l = \SI{20}{\micro\meter}$), the single-harmonic estimate $\sigma_\mathrm{20\si{\micro\meter}} = \SI{0.499}{\giga\pascal}$ remains below this value, illustrating the limited reliability of a stress estimate based on a single data point. This procedure is repeated for every measured temperature.

Fig.\,\ref{si:fig-f-vs-harm-tensile}\,(b) depicts the resulting length dependence of the tensile stress for all measured temperatures. All temperatures display qualitatively the same trend that was already discussed for room temperature above. First, the tensile stress increases for shorter strings, consistent with the length-dependent stress relaxation reported in Ref.\,\onlinecite{buckleUniversalLengthDependence2021}. Second, the usage of only a single harmonic for the tensile stress determination is prone to error and underestimates the actual value, thus creating the apparent discontinuity between $\SI{50}{\micro\meter}$ and $\SI{70}{\micro\meter}$.

\end{document}